\documentclass[a4paper,12pt,pdftex]{article}
\usepackage{amssymb,amsmath}
\usepackage{bm}%
\usepackage{stmaryrd}
\usepackage{mathrsfs} 
\usepackage{tikz,pgfplots}
\usepackage{cite}

\def\beq{\begin{equation}}
\def\eeq{\end{equation}}
\def\bea{\begin{eqnarray}}
\def\eea{\end{eqnarray}}
\def\nn{\nonumber}

\def\Z2{\mathbb{Z}_2^2}
\def\g{\mathfrak{g}}
\newtheorem{lemma}{Lemma}[section]
\newtheorem{prop}[lemma]{Proposition}
\newtheorem{thm}[lemma]{Theorem}

\title{$\cal N$-Extension of double-graded supersymmetric and superconformal quantum mechanics}

\author{N. Aizawa\thanks{{E-mail: {\em aizawa@p.s.osakafu-u.ac.jp}}}, K. Amakawa, S. Doi
\\[10pt]
Department of Physical Science, Osaka Prefecture University, \\
Nakamozu Campus, Sakai, Osaka 599-8531, Japan}
\begin{document}
\maketitle
\thispagestyle{empty}

\vfill
\begin{abstract}
In the recent paper\cite{BruDup}, Bruce and Duplij introduced a 
double-graded version of supersymmetric quantum mechanics (SQM).
It is an extension  of  Lie superalgebraic nature of ${\cal N}=1$ SQM to a $\Z2$-graded superalgebra. 
In this work, we propose an extension of Bruce-Duplij model to higher values of $\cal N.$ 
Furthermore, it is shown that our construction of double-graded SQM is a special case of 
the method which converts a given Lie superalgebra to a $\Z2$-graded superalgebra. 
By employing this method one may convert a model of superconformal mechanics to its double-graded version. 
The simplest example of ${\cal N}=1$ double-graded superconformal mechanics is studied in some detail.  
%the construction of $\Z2$-graded superalgebra based on an ordinary Lie superalgebra. 
%This allows us to consider a double-graded version of superconformal mechanics. The simplest example of it is studied in some detail. 
\end{abstract}

%%%%%%%%%%%%%%%%%%%%%%%%%%%%%%%%%%%%%%%%%%%%%%%%%%%%%%%%%%%%%%%%%%%%%%%%%%%%%%%%%%%%%%
\clearpage
\setcounter{page}{1}

\section{Introduction}

 In the recent paper \cite{BruDup}, Bruce and Duplij introduced a double-graded version of supersymmetric quantum mechanics (SQM). It is a model of quantum mechanical Hamiltonian $H$ which can be factorized in two distinct way, i.e., 
$ H = Q_{01}^2 = Q_{10}^2.$ 
The difference from the standard ${\cal N}=2$ SQM is that two supercharges $Q_{01}$ and $ Q_{10}$ have different degree. It  means that they close in commutator, instead of anticommutator, in a central element $Z.$ More explicitly they satisfy the relation $[Q_{01}, Q_{10}] =Z.$ 
Thus the algebra spanned by $ H, Q_{01}, Q_{10}$ and $Z$ is not a superalgebra, 
but a $\Z2$-graded superalgebra \cite{Ree,rw1,rw2,sch}. 
Each subset $ H, Q_{01}$ and $ H, Q_{10}$ forms a standard SQM. 
Thus the whole system of the double-graded SQM can be regarded as a doubling of ${\cal N}=1$ SQM closed in a $\Z2$-graded superalgebra. 
It is elucidated in \cite{BruDup} that this system is different from the generalizations of SQM found in the literatures. 

  The purpose of the present work is to extend the Bruce-Duplij model to higher values of $\cal N $ and also consider double-graded version of superconformal mechanics (SCM). 
This is done by a mapping from a Lie superalgebra to a $\Z2$-graded superalgebra.  
That is, we present a method in which if a matrix representation of the given Lie superalgebra satisfies a certain condition, one may always have a set of matrices which forms a $\Z2$-graded superalgebra.  We shall see that the models of SQM with even ${\cal N} \geq 2$ by Akulov and Kudinov \cite{AkuKud} satisfies the condition so that they produce a  double-graded SQM with higher $\cal N.$ The Bruce-Duplij model is revisited from the ${\cal N} = 2$ extension. This shed a new light on the Bruce-Duplij model and give a new perspective to the  double-graded extensions of SQM. Indeed, we shall also see that many models of SCM satisfy the condition. 
By this fact, one may construct various models of double-graded SCM. 
  
%We show that one may construct models of double-graded SQM with even $\cal N $ based on the model of SQM by Akulov and Kudinov \cite{AkuKud}.  
%Furthermore, we present a method constructing a $ \Z2$-graded superalgebra for a given Lie superalgebra. 
%If a matrix representation of the given Lie superalgebra satisfies a certain condition, one may always have a set of matrices which forms a $\Z2$-graded superalgebra. 
%This method allows us to consider a double-graded version of SCM.  
%Applying this method to a model of SCM, one may  obtain a model of double-graded SCM.

The present work is motivated by recent renewed interest in color superalgebras \cite{Ree,rw1,rw2,sch}. The color superalgebras  is an attempt to generalize Lie superalgebras which was introduced about a half century ago. 
Recall that Lie superalgebras are defined by introducing a $\mathbb{Z}_2$-grading into Lie algebras. 
The basic idea of color superalgebras is to replace the grading group $\mathbb{Z}_2$ with more general abelian groups. The direct product $\Z2 = \mathbb{Z}_2 \times \mathbb{Z}_2 $ is the simplest non-trivial example. 
Since its introduction, 
algebraic and geometric aspects of color superalgebras/supergroups and `higher graded' version of  supermanifolds have been continuously studied till today. 
The reader may refer the references, e.g. in \cite{Bruce,NaPsiJs2}.

On the other hand, physical implication of color superalgebras are not  clear even today. 
There exist some early works discussing on possible connection between color superalgebras and various physical problems such as supergravity, string theory, quasispin formalism and so on   \cite{LuRi,vas,jyw,zhe,LR,Toro1,Toro2}. 
Recently, color superalgebras started appearing again in physics literatures.  
We mention the works on modifying the spacetime symmetry \cite{tol2,Bruce} and works in connection with parastatistics \cite{tol,StoVDJ}. Especially, a modification of four dimensional super-Minkowski spacetime by inclusion of $\mathbb{Z}_2^n$-grading is investigated in \cite{Bruce} and it gives a basis of the double-graded SQM discussed in \cite{BruDup}. 
It is also remarkable that $ \Z2$-graded superalgebras appear as symmetries of partial differential equations. In \cite{AKTT1,AKTT2} it is demonstrated that symmetries of wave  equations of nonrelativistic quantum mechanics for fermions are given by a $\Z2$-graded extension of Schr\"odinger algebra. 
Furthermore, it is suggested that mixed symmetry tensors over  $\mathbb{Z}_2^n$-graded manifolds will have some connection to the double field theory \cite{BruIba}. 

 We think that these works suggest that there exist more places in physics where color superalgebras play significant roles. 
The results presented here will give an another example of intimate relation between physics and color superalgebras and will provide an tool for physical applications of $\Z2$-graded ones.  
 
 The plan of this paper is as follows:  
We start with giving a definition of $\Z2$-graded superalgebras. It follows by an example of matrix $\Z2$-graded superalgebra which is a generalization of the one given in \cite{BruDup} and will be used to construct $\cal N$-extension of double-graded SQM. 
In \S \ref{SEC:NSQM} a double-graded SQM with higher $\cal N$ is constructed by mapping the $\cal N$-extended SQM of Akulov and Kudinov to a $\Z2$-graded version, then $ {\cal N}=2$ case is studied in some detail. 
%In \S \ref{SEC:NSQM} a double-graded extension of the model of SQM by Akulov and Kudinov is presented and $ {\cal N}=2$ case is studied in some detail. 
In \S \ref{SEC:Z2toZ22} a general method  of converting a Lie superalgebra to $\Z2$-graded one is presented. 
This is based on matrix representations and the matrices in \S \ref{SEC:matrix} are used. 
The model of double-graded SQM in \S \ref{SEC:NSQM} is a special case of this construction. 
The construction is applied to a model of ${\cal N}=1$ SCM and we obtain a double-graded version of SCM in \S \ref{SEC:DGSCM}. Properties of the model is studied in some detail.

%%%%%%%%%%%%%%%%%%%%%%%%%%%%%%%%%%%%%%%%%%%%%%%%%%%%%%%%%%%%%%%%%%%%%%%%
%
%
%   Preliminaries
%
%
%%%%%%%%%%%%%%%%%%%%%%%%%%%%%%%%%%%%%%%%%%%%%%%%%%%%%%%%%%%%%%%%%%%%%%%%%
%
\section{Preliminaries}

\subsection{$\Z2$-graded superalgebras} \label{SEC:Pre}

Here we give the definition of a $\Z2$-graded  superalgebra \cite{rw1,rw2}. 
Let $ \g $ be a vector space over $\mathbb{C}$ and $ \bm{a} = (a_1, a_2)$ an element of  $\Z2$. 
Suppose that $ \g $ is a direct sum of graded components:
\begin{equation}
   \g = \bigoplus_{\bm{a}} \g_{\bm{a}} = \g_{(0,0)} \oplus \g_{(0,1)} \oplus \g_{(1,0)} \oplus \g_{(1,1)}.
\end{equation}
In what follows, we denote homogeneous elements of $ \g_{\bm{a}} $ as $ X_{\bm{a}}, Y_{\bm{a}},
Z_{\bm{a}}$ and refer $ \bm{a}$ to the degree of $ X_{\bm{a}}. $  
If $\g$ admits a bilinear operation (the general Lie bracket), denoted by $ \llbracket \cdot, \cdot \rrbracket, $ 
satisfying the identities
\begin{align}
  & \llbracket X_{\bm{a}}, Y_{\bm{b}} \rrbracket \in \g_{\bm{a}+\bm{b}}
  \\[3pt]
  & \llbracket X_{\bm{a}}, Y_{\bm{b}} \rrbracket = -(-1)^{\bm{a}\cdot \bm{b}} \llbracket Y_{\bm{b}}, X_{\bm{a}} \rrbracket,
  \\[3pt]
  & (-1)^{\bm{a}\cdot\bm{c}} \llbracket X_{\bm{a}}, \llbracket Y_{\bm{b}}, Z_{\bm{c}} \rrbracket \rrbracket
    + (-1)^{\bm{b}\cdot\bm{a}} \llbracket Y_{\bm{b}}, \llbracket Z_{\bm{c}}, X_{\bm{a}} \rrbracket \rrbracket
    + (-1)^{\bm{c}\cdot\bm{b}} \llbracket Z_{\bm{c}}, \llbracket X_{\bm{a}}, Y_{\bm{b}} \rrbracket
\rrbracket =0, 
    \label{gradedJ}
\end{align}
where
\begin{equation}
  \bm{a} + \bm{b} = (a_1+b_1, a_2+b_2) \in {\mathbb Z}_2 \times {\mathbb Z}_2, \qquad \bm{a}\cdot \bm{b} = a_1 b_1 + a_2 b_2,
\end{equation}
then $\g$ is referred to as a $\Z2$-graded superalgebra. 

We take $\g$ to be contained in its enveloping algebra, via the identification
\begin{equation}
 \llbracket X_{\bm{a}}, Y_{\bm{b}} \rrbracket =  X_{\bm{a}} Y_{\bm{b}} - (-1)^{\bm{a}\cdot \bm{b}}
Y_{\bm{b}} X_{\bm{a}}, \label{gradedcom}
\end{equation}
where an expression such as $ X_{\bm{a}} Y_{\bm{b}}$ is understood to denote the associative product
on the enveloping algebra. 
In other words, by definition, in the enveloping algebra the general Lie bracket $ \llbracket \cdot, \cdot
\rrbracket $ for homogeneous elements coincides with either a commutator or anticommutator. 

This is a natural generalization of Lie superalgebra which is defined with a $\mathbb{Z}_2$-graded structure:
\begin{equation}
  \g = \g_{(0)} \oplus \g_{(1)}
\end{equation}
with 
\begin{equation}
  \bm{a} + \bm{b} = (a+b), \qquad \bm{a} \cdot \bm{b} = ab.
\end{equation}
It should be noted that  $ \g_{(0,0)} \oplus \g_{(0,1)} $ and $ \g_{(0,0)} \oplus \g_{(1,0)} $ are
subalgebras of $\g$ (with $\Z2$-grading).

%%%%%%%%%%%%%%%%%%%%%%%%%%%%%%%%%%%%%%%%%%%%%%%%%%%%%%%%%%%%%%%%%%%%%%%%%%%%%%
%
\subsection{A matrix $\Z2$-graded superalgebra} \label{SEC:matrix}

 In the construction of the double-graded SQM in \cite{BruDup}, a matrix $\Z2$-graded superalgebra, generated by `higher' Pauli matrices, plays the fundamental role. 
Here we give an extended version of the matrix $\Z2$-graded superalgebra which will be used to construct an $\cal N$-extension of the double-graded SQM. 

For a given positive integer $n$ we introduce the ${\cal N} = 2n$  Hermitian block-antidiagonal matrices subject to the anticommutation relations:
\begin{equation}
  \{ \gamma_{I}, \gamma_J \} := \gamma_{I} \gamma_{J} + \gamma_{J}\gamma_{I} = 2\delta_{IJ}\, \mathbb{I}_{2^n}, \qquad 
  \gamma_I^{\dagger} = \gamma_I 
  \label{gamma-def}
\end{equation}
where $ I, J $ run from $1$ to ${\cal N} $ and  
$ \mathbb{I}_m $ denotes the $m \times m$ identity matrix.  
Such matrices are given by a irreducible representation of the complex Clifford algebra \cite{Okubo,CaRoTo}. 
We mainly work on an alternative choice of the basis of $\gamma$-matrices:
\begin{equation}
  \gamma_a^{\pm} = \frac{1}{2} (\gamma_{2a-1} \pm i \gamma_{2a}), \quad a = 1, 2, \dots, n. 
\end{equation}
In this basis the relation \eqref{gamma-def} reads as follows:
\begin{equation}
   \{ \gamma_a^{\pm}, \gamma_b^{\pm} \} = 0, \qquad 
   \{ \gamma_a^+, \gamma_b^- \} = \delta_{ab}\, \mathbb{I}_{2^n}.
\end{equation}
We also consider $n$ Hermitian block-diagonal matrices given by a product of $\gamma_I$'s:
\begin{equation}
   \Gamma_a = i^a \gamma_1 \gamma_2 \dots \gamma_{2a}, \quad 
   a = 1, 2, \dots, n. 
\end{equation}
It is then immediate to verify that
\begin{equation}
   \Gamma_a^2 = \mathbb{I}_{2^n}, \qquad [\Gamma_a, \Gamma_b] = 0
\end{equation}
and
\begin{equation}
  [\gamma_k^{\pm}, \Gamma_a] = 0 \quad (k > a), \qquad
  \{\gamma_k^{\pm}, \Gamma_a \} = 0 \quad (k \leq a)
\end{equation}
where $ [A, B] := AB-BA $ denotes a commutator as usual.

Now we give a generalization of higher Pauli matrices. Define $2{\cal N}+1$ Hermitian matrices
\begin{equation}
  F = \begin{pmatrix}
     0 & \Gamma_n \\ \Gamma_n & 0
  \end{pmatrix},
  \qquad
  \Sigma_a^{\pm} = \begin{pmatrix}
     \gamma_a^{\pm} & 0 \\ 0 & \gamma_a^{\pm}
  \end{pmatrix},
  \qquad
  \alpha_a^{\pm} = i\Sigma_a^{\pm} F. 
  \label{BDmatrix}
\end{equation}
Then $ F $ is idempotent and $ \Sigma_a^{\pm}, \alpha_a^{\pm}$ are nilpotent: 
$ F^2 = \mathbb{I}_{2^{n+1}}, (\Sigma_a^{\pm})^2 = (\alpha_a^{\pm})^2 = 0. $ 
One may assign the $\Z2$-degree to the matrices in the following way:
\begin{equation}
  \deg(\mathbb{I}_{2^{n+1}}) = (0,0), \quad 
  \deg(\Sigma_a^{\pm}) = (0,1), \quad
  \deg(\alpha_a^{\pm}) = (1,0), \quad
  \deg(F) = (1,1) 
  \label{Z2degree}
\end{equation}
These matrices have  $4 \times 4$ block structure. 
The position of non-zero entries is determined by the  $\Z2$-degree as follows
\begin{equation}
  \begin{pmatrix}
     (0,0) & (0,1) & (1,1) & (1,0) \\
     (0,1) & (0,0) & (1,0) & (1,1) \\
     (1,1) & (1,0) & (0,0) & (0,1) \\
     (1,0) & (1,1) & (0,1) & (0,0)
  \end{pmatrix}
\end{equation}
We keep this convention in our extension of the double-graded SQM. 
An important fact on the matrices \eqref{BDmatrix} is that they generate a $\Z2$-graded superalgebra.
\begin{prop} \label{Prop:MatrixAlg}
The vector space over $\mathbb{C}$ spanned by the matrices given in \eqref{BDmatrix} and $\mathbb{I}_{2^{n+1}}$ forms a $\Z2$-graded superalgebra. 
Non-vanishing general Lie brackets are (in commutator and anticommutator notation)
\begin{equation}
  \{ \Sigma_a^+, \Sigma_b^- \} = \{ \alpha_a^+, \alpha_b^- \} = \delta_{ab} \mathbb{I}_{2^{n+1}}, \qquad
  [\Sigma_a^{\pm}, \alpha_b^{\mp} ] = i \delta_{ab} F.
\end{equation}
\end{prop}
Proof. One may verify by direct computation that the matrices satisfy the definition $\Z2$-graded superalgebra.  \hfill $\square$

 In fact, the complex vector space spanned by the matrices
 \begin{equation}
    \mathbb{I}_{2^n}, \qquad \gamma_a^{\pm}, \qquad 
    \tilde{\gamma}_a^{\pm} = i \gamma_a^{\pm} \Gamma_n, \qquad  \Gamma_n
 \end{equation}
forms a  $\Z2$-graded superalgebra isomorphic to the one in Proposition \ref{Prop:MatrixAlg}. 
The isomorphism $ \iota $ is given by
\begin{equation}
  \iota(\mathbb{I}_{2^n}) = \mathbb{I}_{2^{n+1}}, \quad 
  \iota(\gamma_a^{\pm}) = \Sigma_a^{\pm}, \quad
  \iota(\tilde{\gamma}_a^{\pm}) = \alpha_a^{\pm}, \quad
  \iota(\Gamma_n) = F.
\end{equation}
The proof of this fact is also straightforward.

%%%%%%%%%%%%%%%%%%%%%%%%%%%%%%%%%%%%%%%%%%%%%%%%%%%%%%%%%%%%%%%%%%%%%%%%
%
%
%   
%
%
%%%%%%%%%%%%%%%%%%%%%%%%%%%%%%%%%%%%%%%%%%%%%%%%%%%%%%%%%%%%%%%%%%%%%%%%%
%
\section{$\cal N$-extension of double-graded SQM} \label{SEC:NSQM}
\setcounter{equation}{0}

In this section, we present an $\cal N$-extension of Bruce-Duplij model. 
It is a double-graded version of $\cal N$-extended SQM by Akulov and Kudinov. 

First, let us recall the Akulov-Kudinov construction of SQM \cite{AkuKud}. 
The $\cal N$ supercharges are defined by the matrices $ \Gamma_a $ as follows:
\begin{equation}
  Q_a^+ = \frac{1}{\sqrt{2}} \gamma_a^+ (p+iW_a^{(n)}(x,\Gamma_1, \dots, \Gamma_n)),
  \qquad
  Q_a^- = (Q_a^+)^{\dagger}. \label{AKsuperC}
\end{equation}
The superpotentials $W^{(n)}$ are defined recursively. 
For instance, if $n=1$, then $ W^{(1)} = w_0(x)$ and one may choose 
$ \gamma_1 = \sigma_1, \gamma_2 = \sigma_2$ which gives $ \Gamma_1 = -\sigma_3.$ 
This case corresponds to  the Witten model of $ {\cal N} = 2$ \cite{Junker}.  
Higher value of $n$ is more involved. 
For $n = 2 $ one may choose
\begin{equation}
   \gamma_1 = \sigma_2 \otimes \sigma_1, \quad 
   \gamma_2 = \sigma_2 \otimes \sigma_2, \quad 
   \gamma_3 = \sigma_2 \otimes \sigma_3, \quad 
   \gamma_4 = \sigma_1 \otimes \mathbb{I}_2 
\end{equation}
and this choice yields $ \Gamma_2 = -\sigma_3 \otimes \mathbb{I}_2.$ 
The superpotential  is given by
\begin{equation}
  W^{(2)}_1 = w_0(x) + \Gamma_2 w_1(x), \quad 
  W^{(2)}_2 = w_0(x) + \Gamma_1 w_1(x),  
  \quad
  w_1(x) = \frac{\partial_x w_0(x)}{2w0(x)}.
\end{equation}
The $n=3$ superpotential reads, e.g.,
\begin{equation}
   W^{(3)}_1 = w_0(x) + \Gamma_2 w_1(x) + \Gamma_3 (w_1(x) + \Gamma_2 w_2(x))
\end{equation}
and $w_0, w_1, w_2$ are functions of a single undetermined function  (see \cite{AkuKud} for more detail). 
 
For all possible values of $n$, the supercharges \eqref{AKsuperC} are nilpotent and the supersymmetric algebra is given by 
\begin{equation}
  \{Q_a^+, Q_b^- \} = \delta_{ab} H, \qquad [H, Q_a^{\pm} ] = 0.
\end{equation}
An important observation is that $\Gamma_n$ anticommutes with the supercharges:
\begin{equation}
  \{ Q_a^{\pm}, \Gamma_n \} = 0, \quad \forall a
\end{equation}

 With this setting of SQM and the matrices in \S \ref{SEC:matrix}, one may extend the double-graded SQM of \cite{BruDup}. 
\begin{thm} \label{Thm:SQM}
The complex vector space spanned by the matrix differential operators
\begin{equation}
   {\cal Q}_a^{\pm} = \mathbb{I}_2 \otimes Q_a^{\pm}, \quad \tilde{\cal Q}_a^{\pm} =  i {\cal Q}_a^{\pm} F, \quad 
   {\cal H} = \mathbb{I}_2 \otimes H, \quad \tilde{\cal H} = {\cal H} F 
   \label{NSUSY}
\end{equation}
forms a $\Z2$-graded superalgebra having the following non-vanishing relations
\begin{equation}
   \{ {\cal Q}_a^+, {\cal Q}_b^- \} = \{ \tilde{\cal Q}_a^+, \tilde{\cal Q}_b^- \} = \delta_{ab} {\cal H}, \qquad
   [ {\cal Q}_a^{\pm}, \tilde{\cal Q}_b^{\mp} ] = i \delta_{ab} \tilde{\cal H}.
   \label{NSUSYrel}
\end{equation}
The assignment of $\Z2$-degree is 
\begin{equation}
   \deg({\cal H}) = (0,0), \quad \deg({\cal Q}_a^{\pm}) = (0,1), \quad
   \deg(\tilde{\cal Q}_a^{\pm}) = (1,0), \quad \deg(\tilde{\cal H}) = (1,1).
\end{equation}
\end{thm}

The theorem is easily proved by direct computation so we omit to include the proof. 
We would like to emphasize one more time that, due to the $\Z2$-grading,  
we take a commutator for ${\cal Q}_a^{\pm}$ and $ \tilde{\cal Q}_b^{\pm}. $

The $\Z2$-graded superalgebra relations in \eqref{NSUSYrel} are identical to the one in \cite{Bruce} and the matrix differential operators \eqref{NSUSY} is a generalization of the ones presented in \cite{BruDup}. Thus the Theorem \ref{Thm:SQM} defines a model of $\cal N$ extended version of double-graded SQM. The Hilbert space of the model is taken to be 
$ \mathscr{H} = L^2(\mathbb{R})\otimes \mathbb{C}^{2^{n+1}}$ and $ {\cal H},  \tilde{\cal H} $ are Hermitian in $\mathscr{H}$ and $ ({\cal Q}_a^{\pm})^{\dagger} = {\cal Q}_a^{\mp}, (\tilde{\cal Q}_a^{\pm})^{\dagger} = \tilde{\cal Q}_a^{\mp}.$ 
The Hilbert space has a vector space decomposition according to the $\Z2$-degree:
\begin{equation}
  \mathscr{H} = \mathscr{H}_{00} \oplus \mathscr{H}_{01} \oplus \mathscr{H}_{10} \oplus \mathscr{H}_{11}.
\end{equation}
It follows readily from \eqref{NSUSYrel} that the eigenvalue of $\cal H$ is not negative. 
It is also easy to see the degeneracy of the excited states. 
Let $ \psi_{ij}(x) \in \mathscr{H}_{ij} $ be a wavefunction of a excited state, namely, 
$ {\cal H} \psi_{ij}(x) = E \psi_{ij}(x) $ with $ E > 0. $ 
Then the wavefunctions in the different subspaces of $ \mathscr{H}$ given by
\begin{equation}
  {\cal Q}_a^{\pm} \psi_{ij} \in \mathscr{H}_{i\; j+1}, \qquad
  \tilde{\cal Q}_a^{\pm} \psi_{ij} \in \mathscr{H}_{i+1\; j}, \qquad
  {\cal Q}_a^{\pm} \tilde{\cal Q}_b^{\pm} \psi_{ij}, \ 
  {\cal Q}_a^{\pm} \tilde{\cal Q}_b^{\mp} \psi_{ij} \in \mathscr{H}_{i+1\; j+1}
\end{equation}
are also the eigenfunctions of $ {\cal H}$ with the same eigenvalue $E,$ 
because $ {\cal Q}_a^{\pm} $ and $ \tilde{\cal Q}_a^{\pm} $ commute with $ \cal H.$

Now let us have a closer look at $n=1\ ({\cal N}=2)$ case.  Recalling that $ \gamma^{\pm} = \sigma_{\pm}$ for $n=1,$ the supercharges \eqref{AKsuperC} are given by
\begin{equation}
  Q^{\pm} = \frac{1}{\sqrt{2}} \sigma_{\pm} (p \pm i W_0(x)) \label{n1SC}
\end{equation}
and Hamiltonian of the standard SQM yields
\begin{equation}
   H = \text{diag }(H_1, H_2), \qquad H_1 = A^{\dagger} A, \quad H_2 = A A^{\dagger}
\end{equation}
with
\begin{equation}
   A = \frac{1}{\sqrt{2}} (ip+W_0), \qquad A^{\dagger} = \frac{1}{\sqrt{2}} (-ip+W_0).
\end{equation}
Since $ \Gamma_1 = -\sigma_3,$ the $\Z2$-graded supercharges \eqref{NSUSY} are written in a way
\begin{equation}
   {\cal Q}^{\pm} = 
   \begin{pmatrix}
      Q^{\pm} & 0 \\ 0 & Q^{\pm}
   \end{pmatrix},
   \qquad
   \tilde{\cal Q}^{\pm} = \pm i 
   \begin{pmatrix}
     0 & Q^{\pm} \\ Q^{\pm} & 0
   \end{pmatrix}  \label{n1Z2SC}
\end{equation}
and we have
\begin{equation}
   {\cal H} = \text{diag}(H_1,H_2,H_1,H_2), \qquad
   \tilde{\cal H} = 
   \begin{pmatrix}
       0 & 0 & -H_1 &0  \\
      0  & 0  & 0  &  H_2 \\
       -H_1 & 0 & 0 & 0  \\
       0 & H_2 & 0 & 0  
   \end{pmatrix}.
\end{equation}

Let us first relate the present model to the Bruce-Duplij model.
From \eqref{n1SC}  we have
\begin{equation}
   Q^+ + Q^- = \frac{1}{\sqrt{2}} (\sigma_1 p - \sigma_2 W_0(x)),
   \quad
   Q^+ - Q^- = \frac{i}{\sqrt{2}} ( \sigma_2 p + \sigma_1 W_0(x)).
\end{equation}
It follows that
\begin{align}
   {\cal Q}^+ + {\cal Q}^- &= 
   \frac{1}{\sqrt{2}} \left(
     \begin{pmatrix}
       \sigma_1 & 0 \\ 0 & \sigma_1
     \end{pmatrix} p
     - 
     \begin{pmatrix}
        \sigma_2 & 0 \\ 0 & \sigma_2
     \end{pmatrix} W_0(x)
   \right),
   \\[3pt]
   \tilde{\cal Q}^+ + \tilde{\cal Q}^- &= 
   -\frac{1}{\sqrt{2}} \left(
     \begin{pmatrix}
        0 & \sigma_2 \\ \sigma_2 & 0
     \end{pmatrix} p
     + 
     \begin{pmatrix}
        0 & \sigma_1 \\ \sigma_1 & 0 
     \end{pmatrix} W_0(x)
   \right).
\end{align}
Therefore we see that $ {\cal Q}^+ + {\cal Q}^- $ and $ \tilde{\cal Q}^+ + \tilde{\cal Q}^- $ correspond to $ Q_{01}$ and $ Q_{10}$ of \cite{BruDup}, respectively.

\begin{figure} 
\begin{center}
   \begin{tikzpicture}
      \node at (0,3.3) {$0$};
      \node at (0,2) {$ \psi_{00}$};
      \node at (0,-2) {$ \psi_{11}$};
      \node at (0,-3.3) {$0$};
      \node at (-2,0) {$\psi_{01}$};
      \node at (-4,0) {$0$};
      \node at (2,0) {$\psi_{10}$};
      \node at (4,0) {$0$};  
      \draw[thick,->] (0,2.3)--(0,3);
      \node at (-0.5,2.6) {${\cal Q}^+ $};
      \node at (0.6,2.6) {$ \tilde{\cal Q}^+ $};
      \draw[thick,->] (0,-2.3)--(0,-3);
      \node at (-0.5,-2.6) {${\cal Q}^+ $};
      \node at (0.6,-2.6) {$ \tilde{\cal Q}^+ $};
      \draw[thick,->] (-2.5,0)--(-3.5,0);      
      \node at (-3,0.4) {$ \tilde{\cal Q}^-$};
      %  00 <--> 01
      \draw[thick,->] (-0.5,1.8)--(-2,0.4);
      \draw[thick,<-] (-0.2,1.7)--(-1.7,0.31);
      \node at (-1.4,1.6) {${\cal Q}^-$};
      \node at (-0.7,0.6) {${\cal Q}^+$};
      \draw[thick,->] (2.5,0)--(3.5,0);      
      \node at (3.1,0.4) {$ {\cal Q}^-$};
      % (01) <--> (11)
      \draw[thick,->] (-2,-0.4)--(-0.5,-1.8);
      \draw[thick,<-] (-1.7,-0.31)--(-0.2,-1.7);
      \node at (-1.4,-1.6) {$\tilde{\cal Q}^+$};
      \node at (-0.7,-0.6) {$\tilde{\cal Q}^-$};
      %  (00) <--> (10)
      \draw[thick,->] (0.5,1.8)--(2,0.4);
      \draw[thick,<-] (0.2,1.7)--(1.7,0.31);
      \node at (1.4,1.6) {$\tilde{\cal Q}^-$};
      \node at (0.7,0.6) {$\tilde{\cal Q}^+$};
      %  (10) <--> (11)
      \draw[thick,->](2,-0.4)--(0.5,-1.8);
      \draw[thick,<-] (1.7,-0.31)--(0.2,-1.7);      
      \node at (1.4,-1.6) {${\cal Q}^+$};
      \node at (0.7,-0.6) {${\cal Q}^-$};
   \end{tikzpicture}
   \caption{Action of $\Z2$-graded supercharges.}  \label{Fig:Qaction}
\end{center}
\end{figure}
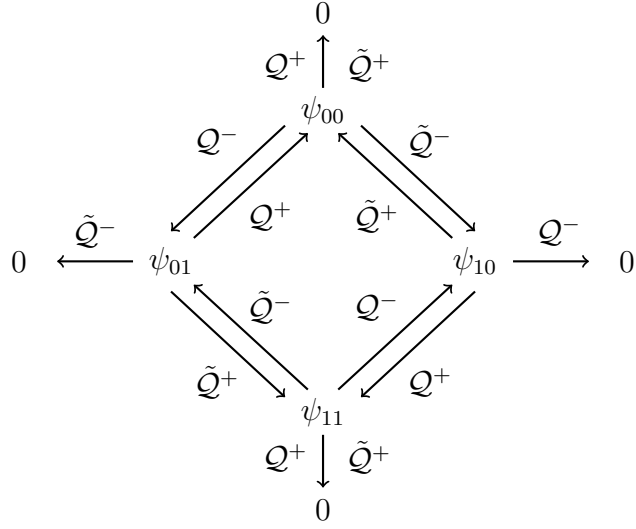

  Next we study the eigenspace of $\cal H$ in some detail. 
Let $\varphi(x)$ be an eigenfunction of $H_1 $ with positive eigenvalue $E$: 
$ H_1 \varphi(x) = E \varphi(x).$ 
We define functions in $\mathscr{H}$ by
\begin{equation}
   \psi_{00}(x) = 
   \begin{pmatrix}
     \varphi(x) \\ 0 \\ 0 \\ 0
   \end{pmatrix},
   \quad
   \psi_{01}(x) = 
   \begin{pmatrix}
     0 \\ A\varphi(x) \\ 0 \\ 0 
   \end{pmatrix},
   \quad
   \psi_{10}(x) = 
   \begin{pmatrix}
    0 \\ 0 \\ 0 \\  A\varphi(x) 
   \end{pmatrix},
   \quad
   \psi_{11}(x) = 
   \begin{pmatrix}
     0 \\ 0 \\\varphi(x) \\ 0 
   \end{pmatrix}
\end{equation}
Assuming that $ \psi_{00}(x) \in \mathscr{H}_{00} $ it is immediate to see that
\begin{equation}
  {\cal H} \psi_{00} = E \psi_{00}, \qquad
  \tilde{\cal H} \psi_{00} = -E \psi_{11}
\end{equation}
so that $ \psi_{11} \in \mathscr{H}_{11}. $ 
The action of the $\Z2$-graded supercharges \eqref{n1Z2SC} is also obtained immediately. 
We list only the non-vanishing action below  (the equality of the equations below is up to a numerical factor):
\begin{alignat}{2}
  {\cal Q}^- \psi_{00} &= \psi_{01},  &\qquad  \tilde{\cal Q}^- \psi_{00} &= \psi_{10}, \nn\\
  {\cal Q}^+ \psi_{01} &= \psi_{00},  &  \tilde{\cal Q}^+ \psi_{10} &= \psi_{00}, \nn \\
  \tilde{\cal Q}^+ \psi_{01} &= \psi_{11},  &  \tilde{\cal Q}^+ \psi_{10} &= \psi_{11}, \nn \\
  {\cal Q}^- \psi_{11} &= \psi_{10},  &  \tilde{\cal Q}^- \psi_{11} &= \psi_{01}. 
\end{alignat}
Therefore $ \psi_{ij}(x) \in \mathscr{H}_{ij}$ for all $i,j$ is seen from these relations. 
The action of the $\Z2$-graded supercharges is summarized in Figure \ref{Fig:Qaction}. 
We see that each positive energy levels of $ \cal H$ are four-fold degenerate. 

  Now we proceed to determine the zero energy ground state of the Hamiltonian $\cal H$ which is given by 
\begin{equation}
   {\cal Q}^{\pm} \Psi(x) = \tilde{\cal Q}^{\pm} \Psi(x) = 0.
\end{equation}
Writing $\Psi(x)$ in components as $ \Psi = (\psi_1(x), \psi_2(x), \psi_3(x), \psi_4(x))^T,$
annihilation by $ {\cal Q}^+ $ and $ \tilde{\cal Q}^+$ gives the relation 
$ A^{\dagger} \psi_2 = A^{\dagger} \psi_4 = 0, $ while  annihilation by $ {\cal Q}^- $ and $\tilde{\cal Q}^- $ gives $ A \psi_1 = A \psi_3 = 0.$ These are the same relations as ${\cal N}=1$ case discussed in \cite{BruDup}. Therefore, the ground state is either non-existent or two-fold degenerate and belongs to $ \mathscr{H}_{00} \oplus \mathscr{H}_{11}$ or $ \mathscr{H}_{01} \oplus \mathscr{H}_{10}. $

%%%%%%%%%%%%%%%%%%%%%%%%%%%%%%%%%%%%%%%%%%%%%%%%%%%%%%%%%%%%%%%%%%%%%%%%
%
%
%    From superalgebra to Z22-graded superalgebra
%
%
%%%%%%%%%%%%%%%%%%%%%%%%%%%%%%%%%%%%%%%%%%%%%%%%%%%%%%%%%%%%%%%%%%%%%%%%%
%
\section{From superalgebra to $\Z2$-graded superalgebra} \label{SEC:Z2toZ22}
\setcounter{equation}{0}

Let $\mathfrak{s}$ be an ordinary Lie superalgebra ($\mathbb{Z}_2$-graded Lie algebra) spanned by the elements $ T_i^a $ with $ a \in \mathbb{Z}_2 = \{0, 1\}. $ 
The defining relations may be written as
\begin{equation}
  [T_i^0, T_j^0] = if_{ij}^k T_k^0, \qquad
  [T_i^0, T_j^1] = i h_{ij}^k T_k^1, \qquad
  \{ T_i^1, T_j^1 \} = g_{ij}^k T_k^0  \label{s-def}
\end{equation}  
where the summation over the repeated indices is understood. 

Suppose that we have a representation of $\mathfrak{s}$ in which odd (degree $1$) elements are represented by block-antidiagonal Hermitian matrix of dimensions $2m \times 2m.$ 
Suppose further that there exists a Hermitian block-diagonal matrix $\Gamma$ of the same dimension which satisfies the relations
\begin{equation}
   \{ \Gamma, T_i^1 \} = 0, \qquad \Gamma^2 = \mathbb{I}_{2m} \label{GammaCond}
\end{equation}
where $ T_i^1$ denotes the matrix representation of $\mathfrak{s}$ (slight abuse of notation). 
It then follows that $[ \Gamma, T_i^0] = 0. $ 

  Now we define two sets of Hermitian matrices: the first one is given by
\begin{equation}
   T_i^a, \qquad \tilde{T}_i^1 = i T_i^1 \Gamma, \qquad \tilde{T}_i^0 = T_i^0 \Gamma 
   \label{Set1}
\end{equation}
and the second one is
\begin{equation}
  {\cal T}_i^a = \mathbb{I}_2 \otimes T_i^a, \qquad 
  \tilde{\cal T}_i^a = \sigma_1 \otimes \tilde{T}_i^a.  \label{Set2}
\end{equation}
With these setting we have the followings:
\begin{prop} \label{Prop:Set1}
   Let $\hat{\mathfrak{s}}$ be the complex vector space spanned by the matrices \eqref{Set1}. 
  By the assignment of the $\Z2$-degree 
  \begin{equation}
      \deg(T_i^0) = (0,0), \quad \deg(T_i^1) = (0,1), \quad
      \deg(\tilde{T}_i^1) = (1,0), \quad \deg(\tilde{T}_i^0) = (1,1), 
  \end{equation}
  $\hat{\mathfrak{s}}$ forms a $\Z2$-graded superalgebra. The defining relations of $\hat{\mathfrak{s}}$ are given as follows:
  \begin{alignat}{2}
    [T_i^0, T_j^0] &= i f_{ij}^k T_k^0, & \qquad [T_i^0, T_j^1] &= i h_{ij}^k T_k^1,
    \nonumber
    \\[3pt]
    [T_i^0, \tilde{T}_j^1] &= i h_{ij}^k \tilde{T}_k^1,   & [T_i^0, \tilde{T}_j^0] &= i f_{ij}^k \tilde{T}_k^0,
    \nonumber
    \\[3pt]
    \{ T_i^1, T_j^1 \} &= \{ \tilde{T}_i^1, \tilde{T}_j^1 \} = g_{ij}^k T_k^0, 
    &
    [\tilde{T}_i^0, \tilde{T}_j^0] &= i f_{ij}^k T_k^0,
    \nonumber 
    \\[3pt]
    [ T_i^1, \tilde{T}_j^1 ] &= i g_{ij}^k \tilde{T}_k^0, & 
    \{ \tilde{T}_i^0, T_j^1 \} &= - h_{ij}^k \tilde{T}_k^1,
    \nonumber
    \\
    \{ \tilde{T}_i^0, \tilde{T}_j^1 \} &= h_{ij}^k T_k^1.  \label{Z22relations}
  \end{alignat}
\end{prop}

\begin{prop} \label{Prop:Set2}
 The complex vector space spanned by the matrices \eqref{Set2} forms a $\Z2$-graded superalgebra which is isomorphic to $\hat{\mathfrak{s}}$. 
The isomorphism $\rho$ is given by 
$ \rho({\cal T}_i^a) = T_i^a, \ \rho(\tilde{\cal T}_i^a) = \tilde{T}_i^a. $ 
The $\Z2$-degree for the matrices is obvious from the isomorphism. 
\end{prop}

Theorem \ref{Thm:SQM} follows immediately  from Proposition \ref{Prop:Set2} by taking $ \mathfrak{s} $ as the supersymmetric algebra with the supercharges \eqref{AKsuperC}. 
If we take a  model of SCM satisfying the condition \eqref{GammaCond}, 
Proposition \ref{Prop:Set2} gives us a model of $\Z2$-graded version of SCM. 
We shall discuss it in the next section. 

  The proof of the propositions is not difficult. 
The relations \eqref{Z22relations} are verified by direct computation using \eqref{s-def} and \eqref{GammaCond}. The matrices \eqref{Set2} satisfy the same relations as \eqref{Z22relations} which is also easily seen. 
Therefore two propositions are proved by verifying the $\Z2$-graded Jacobi identities for $\hat{\mathfrak{s}}. $  
The $\Z2$-graded Jacobi identities for $\hat{\mathfrak{s}}$ are reduced to the ordinary Jacobi identities for the superalgebra $\mathfrak{s}.$ 
To see this we change the notations for $\mathfrak{s}$ and $ \hat{\mathfrak{s}} $ in the following way. 
The $\mathbb{Z}_2$ label $a$ of an element $T_i^a$ of $ \mathfrak{s}$ is replaced with two components vector $ \bm{a} = (0,a),$ i.e., $ T_i^{\bm{a}} = T_i^a. $ 
This allows us to realize the Lie bracket of superalgebras in the same way as \eqref{gradedcom}:
\begin{equation}
  [T_i^{\bm{a}}, T_j^{\bm{b}}]_{\pm} = T_i^{\bm{a}} T_j^{\bm{b}} - 
  (-1)^{\bm{a}\cdot \bm{b}} \, T_j^{\bm{b}} T_i^{\bm{a}}
\end{equation}
where we denote the $\mathbb{Z}_2$-graded Lie bracket of superalgebras by $ [ \cdot, \cdot ]_{\pm}$ in order to make a distinction from the $\Z2$-graded Lie bracket $ \llbracket \cdot, \cdot \rrbracket $  of $\Z2$-graded superalgebras. 
Then the relation \eqref{s-def} may be written in a unified form:
\begin{equation}
   [T_i^{\bm{a}}, T_j^{\bm{b}}]_{\pm} = c_{ij}^k  T_k^{\bm{a}+\bm{b}}
\end{equation}
%where $ [ \cdot, \cdot ]_{\pm}$ is the $\mathbb{Z}_2$-graded Lie bracket realized by commutator or anticommutator:
%\begin{equation}
%  [T_i^{\bm{a}}, T_j^{\bm{b}}]_{\pm} = T_i^{\bm{a}} T_j^{\bm{b}} - 
%  (-1)^{\bm{a}\cdot \bm{b}} \, T_j^{\bm{b}} T_i^{\bm{a}}.
%\end{equation}
Next we unify the matrices $ \mathbb{I}_{2m}$ and $ \Gamma $  in a single expression:
\begin{equation}
   G^{\bm{\alpha}} = 
   \begin{cases}
      \mathbb{I}_{2m} & \bm{\alpha} = (0,0) \\[7pt]
      \Gamma          & \bm{\alpha} = (1,1) 
   \end{cases}
\end{equation}
Note that $ G^{\bm{\alpha}} $ is not defined for $ \bm{\alpha} = (0,1), (1,0). $ 
Then we see the relations
\begin{equation}
   G^{\bm{\alpha}} G^{\bm{\beta}} = G^{\bm{\alpha}+\bm{\beta}}, \qquad
   T_i^{\bm{a}} G^{\bm{\alpha}} = (-1)^{\bm{a}\cdot \bm{\alpha}} G^{\bm{\alpha}} T_i^{\bm{a}}
\end{equation}
where vector sum and the inner product are computed in $\text{mod}\ 2.$ 
With these notations any elements of $\hat{\mathfrak{s}}$ may be expressed as
$ X_i^{\bm{a}+\bm{\alpha}} = i^a T_i^{\bm{a}} G^{\bm{\alpha}}.$ 
The factor $i^a$ does not play any role in the computation of Jacobi identity so that we omit the factor in the following computation. 

The general Lie bracket, realized as in \eqref{gradedcom}, for $\hat{\mathfrak{s}} $  yields
\begin{equation}
  \llbracket X_i^{\bm{a}+\bm{\alpha}}, X_j^{\bm{b}+\bm{\beta}} \rrbracket =  (-1)^{\bm{b}\cdot \bm{\alpha}} 
  [T_i^{\bm{a}}, T_j^{\bm{b}} ]_{\pm}\, G^{\bm{\alpha}+\bm{\beta}}.
\end{equation}
In this computation we used the fact that $\bm{\alpha}\cdot \bm{\beta} = 0\ (\text{mod}\ 2).$ 
We further see that 
\begin{equation}
  \llbracket X_i^{\bm{a}+\bm{\alpha}}, \llbracket X_j^{\bm{b}+\bm{\beta}}, X_k^{\bm{c}+\bm{\gamma}} \rrbracket\, \rrbracket 
  = 
  (-1)^{\bm{a}\cdot \bm{\gamma} + \bm{b}\cdot \bm{\alpha} + \bm{c}\cdot \bm{\beta}} 
  (-1)^{\bm{a}\cdot\bm{c}}
  [\, T_i^{\bm{a}}, [T_j^{\bm{b}}, T_k^{\bm{c}}]_{\pm}\, ]_{\pm} G^{\bm{\alpha}+\bm{\beta}+\bm{\gamma}}.
\end{equation}
Using this relation one may see  that the $\Z2$-graded Jacobi identity \eqref{gradedJ} is reduced to that for a Lie superalgebra.  Thus the propositions are proved.

Before closing this section, some remarks are in order.  
In Proposition \ref{Prop:Set1} the direct sum of $ \mathfrak{s} \oplus \langle\, \Gamma \,\rangle $ is considered. Then $\hat{\mathfrak{s}}$ is realized in the enveloping algebra of the direct sum. 
It is widely known that there are many examples of $\Z2$-graded superalgebra which are realized in the enveloping algebra of a Lie superalgebra, see e.g.  \cite{AKTT1,AKTT2,NAJS,NaPsiJs,NaPsiJs2}. 
While in Proposition \ref{Prop:Set2} tensor product of $\mathfrak{s}$ and Clifford algebra gives a $\Z2$-graded superalgebra. 
In fact, some examples of such mappings that convert Lie superalgebras to $\Z2$-graded superalgebras are known in the  early works \cite{rw1,rw2,sch} and in the recent study \cite{NA2}. It implies that there exist various way mapping Lie superalgebras to $\Z2$-graded ones. The mapping presented in Proposition \ref{Prop:Set2} is the one suitable for SQM. 
%This is an example of the way obtaining $\Z2$-graded superalgebras discussed in the early works \cite{rw1,rw2,sch} and in the recent study \cite{NA2}. 

%%%%%%%%%%%%%%%%%%%%%%%%%%%%%%%%%%%%%%%%%%%%%%%%%%%%%%%%%%%%%%%%%%%%%%%%
%
%
%   Double-graded superconformal mechanics
%
%
%%%%%%%%%%%%%%%%%%%%%%%%%%%%%%%%%%%%%%%%%%%%%%%%%%%%%%%%%%%%%%%%%%%%%%%%%
%
\section{Double-graded superconformal mechanics} \label{SEC:DGSCM}
\setcounter{equation}{0}

As shown in \S \ref{SEC:Z2toZ22}, any Lie superalgebra satisfying the condition \eqref{GammaCond} can be promoted to a $\Z2$-graded superalgebra. 
If one starts with a matrix differential operator realization of a superconformal algebra, i.e. a model of SCM, then one may obtain a double-graded SCM. Many models of SCM have been known. 
The readers may refer the good reviews, e.g. \cite{FedIvaLec,Pa,Oka}. 
Some of the models, e.g. the ones in \cite{AKT,ACKT}, satisfy the condition \eqref{GammaCond} so that we may have the double-graded SCM of ${\cal N}=2, 4, 8$ and so on. 
 
In this section we analyse the simplest example of double-graded SCM obtained from $osp(1|2)$ superconformal algebra. 
Let us consider the following realization of $osp(1|2)$ which is a ${\cal N}=1$ SCM:
\begin{alignat}{2}
  Q &= \frac{1}{\sqrt{2}} \left( \sigma_1 p - \sigma_2 \frac{\beta}{x} \right), 
  & \qquad 
  S &= \frac{x}{\sqrt{2}} \sigma_1, 
  \nn \\
  H &= \frac{1}{2} \left( p^2 + \frac{\beta^2}{x^2} \right) \mathbb{I}_2 + \frac{\beta}{2x^2}\sigma_3,
  & D &= -\frac{1}{4} \{ x, p \}\, \mathbb{I}_2, \qquad 
    K = \frac{x^2}{2}\, \mathbb{I}_2 
    \label{ospSCM}
\end{alignat}
where $ \beta $ is a coupling constant. 
We remark that $Q$ and $ S$ are the same as the one discussed in \cite{ACKT} (see eq.(34) of \cite{ACKT}).  The non-vanishing relations of $osp(1|2)$ read as follows:
\begin{alignat}{3}
 [D, K] &=iK, & \qquad [H, K] &= 2iD, & \qquad [D, H] &= -iH, 
 \nn \\
 \{Q, Q\} &= 2H, & \{S, S\} &= 2K, & \{Q, S\} &=-2D,
 \nn \\
 [D, Q] &=-\frac{i}{2} Q, & [D, S] &= \frac{i}{2} S, & [Q, K]&= -iS, 
 \nn \\
 [S, H] &= iQ.
\end{alignat}
One may immediately see that $\Gamma = \sigma_3 $ satisfies the condition \eqref{GammaCond}. 

  Therefore we obtain the matrix realization of the $\Z2$-graded version of $ osp(1|2)$ which gives a double-graded SCM:
\begin{alignat}{4}
  & (0,0) :  & \quad 
   {\cal H} &= 
    \begin{pmatrix}
       H & 0 \\ 0 & H
    \end{pmatrix},
  & \qquad
    {\cal D} &=
     \begin{pmatrix}
        D & 0 \\ 0 & D
     \end{pmatrix},
  & \qquad
    {\cal K} &=
     \begin{pmatrix}
         K & 0 \\ 0 & K
     \end{pmatrix},
  \nn \\
  & (0,1) : &
    {\cal Q} &=
      \begin{pmatrix}
         Q & 0 \\ 0 & Q
      \end{pmatrix},
  & {\cal S} &=
      \begin{pmatrix}
         S & 0 \\ 0 & S
      \end{pmatrix},
  \nn \\
  & (1,0) : &
   \tilde{\cal Q} &= i
    \begin{pmatrix}
       0 & Q\sigma_3 \\ Q \sigma_3 & 0
    \end{pmatrix},
  & \tilde{\cal S} &= i
     \begin{pmatrix}
         0 & S \sigma_3 \\ S \sigma_3 & 0
     \end{pmatrix},
  \nn \\
  & (1,1) :  & \quad 
   \tilde{\cal H} &= 
    \begin{pmatrix}
       0 &  H\sigma_3 \\  H\sigma_3 & 0
    \end{pmatrix},
  & \qquad
    \tilde{\cal D} &=
     \begin{pmatrix}
        0 & D \sigma_3 \\  D \sigma_3 & 0
     \end{pmatrix},
  & \qquad
    \tilde{\cal K} &=
     \begin{pmatrix}
        0 &  K \sigma_3 \\  K \sigma_3 & 0
     \end{pmatrix}.   
     \label{Z2osp12}
\end{alignat}
For the range of $\beta$ being the potential is repulsive, the  Hamiltonian $H$ in \eqref{ospSCM} has continuous spectrum. 
It is known that the eigenfunctions of $H$ with the positive eigenvalue are plane wave normalizable, however, the zero energy state is not even plane wave normalizable \cite{DFF}. 
This property is inherited to the Hamiltonian $\cal H$ of the double-graded SCM \eqref{Z2osp12}.  
In order to analyse the syetem \eqref{Z2osp12} we follow the standard prescription of conformal mechanics. That is, the eigenspace of $H$ is not taken as the Hilbert space of the theory. Instead, the eigenspace of an operator which is a  linear combination of $H$ and $K$ is chosen as the Hilbert space. We thus introduce the following operators:
%
%The Hamiltonian $\cal H$ has no bound states and furthermore the ground state is not normalizable \cite{DFF}. 
%Following the standard procedure of conformal mechanics we define the operators
\begin{alignat}{2}
  {\cal R} &= {\cal H} + {\cal K}, & \qquad 
  {\cal L}_{\pm} &= \frac{1}{2}({\cal K} - {\cal H}) \pm i {\cal D},
  \nn \\
  \mathtt{a} &= {\cal S} + i {\cal Q}, & 
  \mathtt{a}^{\dagger} &= {\cal S} - i {\cal Q}
\end{alignat}
and similar ones with tilde. 
Then we have
\begin{alignat}{2}
   {\cal R} &= \frac{1}{2} \{ \mathtt{a}, \mathtt{a}^{\dagger} \} = 
   \frac{1}{2}\{ \tilde{\mathtt{a}}, \tilde{\mathtt{a}}^{\dagger} \},
   \nn \\[3pt]
   {\cal L}_+ &=  \frac{1}{4} \{ \mathtt{a}^{\dagger}, \mathtt{a}^{\dagger} \} = 
   \frac{1}{4}\{ \tilde{\mathtt{a}}^{\dagger}, \tilde{\mathtt{a}}^{\dagger} \},
   & \qquad
   {\cal L}_- &=  \frac{1}{4} \{ \mathtt{a}, \mathtt{a} \} = 
   \frac{1}{4}\{ \tilde{\mathtt{a}}, \tilde{\mathtt{a}} \}
\end{alignat}
and
\begin{equation}
   \tilde{\cal R} = \frac{1}{2i} [ \mathtt{a}, \tilde{\mathtt{a}}^{\dagger} ] = 
   \frac{1}{2i} [ \mathtt{a}^{\dagger}, \tilde{\mathtt{a}} ],
   \qquad
    \tilde{\cal L}_+ =  \frac{1}{4i} [ \mathtt{a}^{\dagger}, \tilde{\mathtt{a}}^{\dagger} ],
   \qquad 
   \tilde{\cal L}_- =  \frac{1}{4i}  [\mathtt{a}, \tilde{\mathtt{a}} ].
\end{equation}
Due to the harmonic potential, the new Hamiltonian $\cal R$ may have bound states and its spectrum is discrete. 
The $\Z2$-graded version of $osp(1|2)$ is a spectrum generating algebra of $\cal R$ which can be seen from the relations
\begin{alignat}{2}
 [{\cal R}, {\cal L}_{\pm}] &= \pm 2 {\cal L}_{\pm}, & \qquad
 [{\cal L}_+, {\cal L}_- ] &= -{\cal R},
 \nn \\
 [ {\cal R}, \mathtt{a}] &= - \mathtt{a}, & 
 [ {\cal R}, \mathtt{a}^{\dagger}] &= \mathtt{a}^{\dagger},
 \nn \\
 [ {\cal R}, \tilde{\mathtt{a}}] &= - \tilde{\mathtt{a}}, & 
 [ {\cal R}, \tilde{\mathtt{a}}^{\dagger}] &= \tilde{\mathtt{a}}^{\dagger}.
\end{alignat}
Furthermore, in the realization \eqref{Z2osp12}, we have the relation
\begin{equation}
  [ \mathtt{a}, \mathtt{a}^{\dagger} ] = [ \tilde{\mathtt{a}}, \tilde{\mathtt{a}}^{\dagger} ] =  
  \mathbb{I}_4 - 2\beta {\cal F}, 
  \qquad
  {\cal F} = 
    \begin{pmatrix}
       \sigma_3 & 0 \\ 0 & \sigma_3
    \end{pmatrix}. 
   \label{DeformedBoson}
\end{equation}
$ \cal F $ is one of the parity operators discussed in \cite{BruDup} and commutes with $ \cal R.$ 
With the aid of \eqref{DeformedBoson} one may write $\cal R$ in the following way:
\begin{equation}
  {\cal R} = \mathtt{a}^{\dagger} \mathtt{a} + \frac{1}{2} (\mathbb{I}_4 -2\beta {\cal F})
           = \tilde{\mathtt{a}}^{\dagger} \tilde{\mathtt{a}} + \frac{1}{2} (\mathbb{I}_4 -2\beta {\cal F}).
\end{equation}
Therefore the ground state of $\cal R$ is defined by
\begin{equation}
      \mathtt{a} \Psi(x) = \tilde{\mathtt{a}} \Psi(x) = 0. \label{GSSCM}
\end{equation}
In components $ \Psi(x) = (\psi_1(x), \psi_2(x), \psi_3(x), \psi_4(x))^T$ the condition \eqref{GSSCM} yields
\begin{equation}
  \left( x + ip + \frac{\beta}{x}\right) \psi_{1,3} = \left( x + ip - \frac{\beta}{x}\right) \psi_{2,4} = 0. 
\end{equation}
Thus the ground state is either
\begin{equation}
   C \begin{pmatrix}
      x^{-\beta} e^{-x^2/2} \\ 0 \\ 0 \\ 0
   \end{pmatrix}
   + 
   C' \begin{pmatrix}
      0 \\ 0 \\ x^{-\beta} e^{-x^2/2} \\ 0    
   \end{pmatrix} 
   \in \mathscr{H}_{00} \oplus \mathscr{H}_{11}
   \label{GS1}
\end{equation}
with the energy $ \frac{1}{2}(1-2\beta) $ or
\begin{equation}
   C \begin{pmatrix}
      0 \\ x^{\beta} e^{-x^2/2} \\ 0 \\ 0 
   \end{pmatrix}
   + 
   C' \begin{pmatrix}
      0 \\ 0 \\ 0 \\ x^{\beta} e^{-x^2/2}
   \end{pmatrix} 
   \in \mathscr{H}_{01} \oplus \mathscr{H}_{10}
   \label{GS2}
\end{equation}
with the energy $ \frac{1}{2}(1+2\beta). $   
The normalizability of the wavefunction of the form $ x^{\pm \beta} e^{-x^2/2}$ and the Hilbert space of the SCM  \eqref{ospSCM} are studied in detail in \cite{ACKT}. 
Depending on the value of $\beta $ the wavefunction $ x^{\pm \beta} e^{-x^2/2}$ is normalizable on the half line $(x > 0)$ or on the full line ($ x \in (-\infty,\infty)$). 
It then follows that the admissible Hilbert space of SCM  \eqref{ospSCM} is separated into three cases depending on $\beta $ (see \S IV and Appendix C of \cite{ACKT} for detail).  

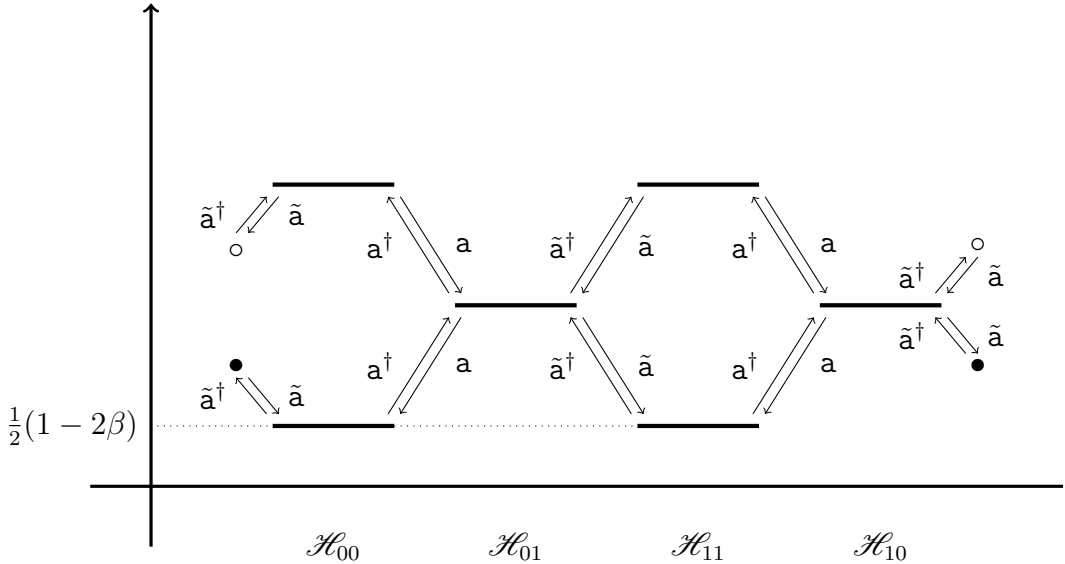
\begin{figure}[h]
\begin{center}
\begin{tikzpicture}[scale=0.8]
\draw[very thick] (-1,0) -- (15, 0);%座標軸 横
\draw[->, very thick] (0,-1) -- (0, 8); % 縦軸
\node (A) at (3,-1)  {$\mathscr{H}_{00}$};%軸下
\node (B) at (6,-1)  {$\mathscr{H}_{01}$};
\node (C) at (9,-1)  {$\mathscr{H}_{11}$};
\node (D) at (12,-1)  {$\mathscr{H}_{10}$};
\draw[ultra thick] (2,1) to  (4,1);%GS
\draw[ultra thick] (8,1) to  (10,1);
\draw[->] (3.9,1.2)--(4.9,2.8)node [midway,left=5pt] {$\mathtt{a}^{\dagger}$};
\draw[<-] (4.1,1.2)--(5.1,2.8)node [midway,right=5pt] {$\mathtt{a}$};
%\draw[<->] (4,1.2) -- (5, 2.8)node [midway,left=5pt] {$\mathtt{a}^{\dagger}$} node [midway,right=5pt] {$\mathtt{a}$};%矢印
\draw[->](9.9,1.2)--(10.9,2.8)node [midway,left=5pt] {$\mathtt{a}^{\dagger}$};
\draw[<-](10.1,1.2)--(11.1,2.8)node [midway,right=5pt] {$\mathtt{a}$};
%\draw[<->] (10,1.2) -- (11, 2.8)node [midway,left=5pt] {$\mathtt{a}^{\dagger}$}node [midway,right=5pt] {$\mathtt{a}$};

\draw[->] (7.9,1.2)--(6.9,2.8)node [midway,left=5pt] {$\tilde{\mathtt{a}}^{\dagger}$};
\draw[<-] (8.1,1.2)--(7.1,2.8)node [midway,right=5pt] {$\tilde{\mathtt{a}}$};
%\draw[<->] (8,1.2) -- (7, 2.8)node [midway,left=5pt] {$\tilde{\mathtt{a}}^{\dagger}$}node [midway,right=5pt] {$\tilde{\mathtt{a}}$};
\draw[ultra thick] (5,3) to  (7,3);%1st ex
\draw[ultra thick] (11,3) to  (13,3);
\draw[->] (4.9,3.2)--(3.9,4.8)node [midway,left=5pt] {$\mathtt{a}^{\dagger}$};
\draw[<-] (5.1,3.2)--(4.1,4.8)node [midway,right=5pt] {$\mathtt{a}$};
%\draw[<->] (5,3.2) -- (4, 4.8)node [midway,left=5pt] {$\mathtt{a}^{\dagger}$} node [midway,right=5pt] {$\mathtt{a}$};%矢印
\draw[->] (10.9,3.2)--(9.9,4.8)node [midway,left=5pt] {$\mathtt{a}^{\dagger}$};
\draw[<-] (11.1,3.2)--(10.1,4.8)node [midway,right=5pt] {$\mathtt{a}$};
%\draw[<->] (11,3.2) -- (10, 4.8)node [midway,left=5pt] {$\mathtt{a}^{\dagger}$}node [midway,right=5pt] {$\mathtt{a}$};
\draw[->] (6.9,3.2)--(7.9,4.8)node [midway,left=5pt] {$\tilde{\mathtt{a}}^{\dagger}$};
\draw[<-] (7.1,3.2)--(8.1,4.8)node [midway,right=5pt] {$\tilde{\mathtt{a}}$};
%\draw[<->] (7,3.2) -- (8, 4.8)node [midway,left=5pt] {$\tilde{\mathtt{a}}^{\dagger}$}node [midway,right=5pt] {$\tilde{\mathtt{a}}$};
\draw[ultra thick] (2,5) to  (4,5);%2nd ex
\draw[ultra thick] (8,5) to  (10,5);
%
%\draw[<-] (2,1.2) -- (1.5, 1.8)node [midway,right=5pt] {$\tilde{\mathtt{a}}$};%矢印
\draw[->] (1.9,1.2) -- (1.4,1.8)node [midway,left=5pt] {$\tilde{\mathtt{a}}^{\dagger}$};
\draw[<-] (2.1,1.2) -- (1.6, 1.8)node [midway,right=5pt] {$\tilde{\mathtt{a}}$};%矢印
\node at (1.4,2) {$\bullet$};
%\draw[->] (1.5,4.2) -- (2, 4.8)node [midway,left=5pt] {$\tilde{\mathtt{a}}^{\dagger}$};%矢印
\draw[->] (1.4,4.2) -- (1.9, 4.8)node [midway,left=5pt] {$\tilde{\mathtt{a}}^{\dagger}$};%矢印
\draw[<-] (1.6,4.2) -- (2.1, 4.8)node [midway,right=5pt] {$\tilde{\mathtt{a}}$};%矢印
\node at (1.4,3.9) {$\circ$};
%
%
%\draw[->] (13.5,2.2) -- (13, 2.8)node [midway,left=5pt] {$\tilde{\mathtt{a}}^{\dagger}$};
\draw[->] (13.4,2.2) -- (12.9, 2.8)node [midway,left=5pt] {$\tilde{\mathtt{a}}^{\dagger}$};
\draw[<-] (13.6,2.2) -- (13.1, 2.8)node [midway,right=5pt] {$\tilde{\mathtt{a}}$};
\node at (13.6,2) {$\bullet$};
%\draw[<-] (13,3.2) -- (13.5, 3.8)node [midway,right=5pt] {$\tilde{\mathtt{a}}$};
\draw[->] (12.9,3.2) -- (13.4, 3.8)node [midway,left=5pt] {$\tilde{\mathtt{a}}^{\dagger}$};
\draw[<-] (13.1,3.2) -- (13.6, 3.8)node [midway,right=5pt] {$\tilde{\mathtt{a}}$};
\node at (13.6,4) {$\circ$};
\draw[dotted] (0,1)--(2,1);
\draw[dotted] (4,1)--(8,1);
\node at (-1.3,1) {$\frac{1}{2}(1-2\beta) $};
\end{tikzpicture}
\end{center}
\caption{Spectrum on \eqref{GS1}.} \label{Fig:SCM}
\end{figure}

  Since the double-graded Hamiltonian $\cal H$ is a two copies of $H,$ the admissible Hilbert space of \eqref{Z2osp12} has  the structure same as \eqref{ospSCM}. 
Namely, either \eqref{GS1} or \eqref{GS2} are normalizable so that the Hilbert space is constructed on one of them by repeated application of $ \mathtt{a}^{\dagger} $ and $ \tilde{\mathtt{a}}^{\dagger}. $ The third possibility is that both \eqref{GS1} and \eqref{GS2} are normalizable so that the Hilbert space is a direct sum of two spaces constructed on \eqref{GS1} and \eqref{GS2}. 
We indicate the spectrum constructed on \eqref{GS1} with the action of creation/annihilation operators in Figure \ref{Fig:SCM}.

%%%%%%%%%%%%%%%%%%%%%%%%%%%%%%%%%%%%%%%%%%%%%%%%%%%%%%%%%%%%%%%%%%%%%%%%
%
%   Concluding remarks
%   
%
%
%%%%%%%%%%%%%%%%%%%%%%%%%%%%%%%%%%%%%%%%%%%%%%%%%%%%%%%%%%%%%%%%%%%%%%%%%
%
\section{Concluding remarks}
\setcounter{equation}{0}

We introduced an ${\cal N}$-extended version of double-graded SQM and SCM. This was done by a matrix differential operator realization of quantum mechanical operators which generate a  $\Z2$-graded superalgebra. Those operators are readily obtained from the standard SQM and SCM. This implies that double-graded SQM and SCM are not rare examples of quantum systems, but quite general ones so that they would have various applications in mathematical physics.  

 We presented a quantum theory based on Lie algebraic method using matrix differential operators. 
Formulation of the corresponding classical theory is a non-trivial problem since we need to deal with variables of degree $(1,1)$ which are not nilpotent and furthermore anticommute with variables of 
degree $(0,1), (1,0).$ This leads us to the notion of  `higher grading' extension of supermanifolds and mathematical theory of them have been developed \cite{Mar,CGP3,CGP4,CoKPo}.  Following those developments  superfields on $\mathbb{Z}_2^n$-graded super-Minkowski spacetime are discussed in \cite{Bruce}.  

  Another interesting problem is a geometric consideration of the double-graded SCM based on $\Z2$-graded supermanifolds. 
A possible approach would be an extension of the work \cite{IvKrLe} where the group manifold of $SO(1,2)$ is taken as a target space, or \cite{MiSt} where more general $N$-dimensional space with torsion is considered. One may also consider an extension of the orbit method to $\Z2$-graded supergroup which would give all possible Hamiltonian systems (see  \cite{Gonera} for the conformal mechanics of de Alfaro-Fubini-Furlan).

%%%%%%%%%%%%%%%%%%%%%%%%%%%%%%%%%%%%%%%%%%%%%%%%%%%%%%%%%%%%%%%%%%%%%%%%
%
%
%
%  References
%
%
%%%%%%%%%%%%%%%%%%%%%%%%%%%%%%%%%%%%%%%%%%%%%%%%%%%%%%%%%%%%%%%%%%%%%%%%%
%

\end{document}